# A Metadata Generation System with Semantic Understanding for Video Retrieval in Film Production

Feilin Han
Department of Film & TV Technology
Beijing Film Academy
Beijing, China
hanfeilin@bfa.edu.cn

Zhaoxu Meng
Department of Film & TV Technology
Beijing Film Academy
Beijing, China
jackie.mzx@gmail.com

**ABSTRACT**

In film production, metadata plays an important role in original raw video indexing and classification within the industrial post-production software. Inspired by deep visual-semantic methods, we propose an automated image information extraction process to extend the diversity of metadata entities for massive large-scale raw video searching and retrieval. In this paper, we introduce the proposed system architecture and modules, integrating semantic annotation models and user-demand-oriented information fusion. We conducted experiments to validate the effectiveness of our system on Film Raw Video Semantic Annotation Dataset (Film-RVSAD) and Slate Board Template Dataset (SBTD), two benchmark datasets built for cinematography-related semantic annotation and slate detection. Experimental results show that the proposed system provides an effective strategy to improve the efficiency of metadata generation and transformation, which is necessary and convenient for collaborative work in the filmmaking process.

## 1 INTRODUCTION

While the industrialization process of film production, how to improve the efficiency of existing filmmaking workflow has attracted much attention thanks to the development of digital media technology. Nowadays, almost all films are shot by digital cameras, which means original shot videos, named raw video, are digitalized. These raw videos are extraordinarily large because they are recorded in RAW format with lossless compression. For a regular commercial film, the originally recorded videos exceed 100TB in general, which is a time-consuming issue for post-production artists to browse all the videos.

To solve this problem, there are several works assigned to improve efficiency. Firstly, when shooting the video, the digital film camera will record basic information of this take and stored them in the metadata, a specific data structure that is designed for video management and digital image engineering. The detailed data structure of metadata depends on the camera manufacturer, however, usually includes typical parameters of cameras, e.g. frame rate, shutter speed, aperture, focus, ISO, and timecode. Another important work in the shooting period is slating. When the director says, 'Action!', the slate board will appear in front of the camera and close with a big sound. It is used to mark the name of this shooting, director, date, as well as the number of the scene, shot, and take, which plays a vital role in post-production editing.

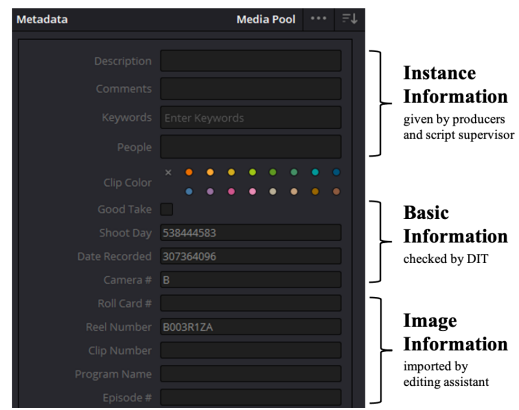

Figure 1: An instance metadata example after being imported into the pose-production software.

Meanwhile, a script supervisor is employed on the film set, who is responsible for recording the cinematography-related information and scene details, such as lens number, camera movement, shot scale, time length, actors, dialogue, simple description of the performance, props, etc., in a detailed and accurate continuity report.

Metadata could be extracted while videos imported into the post-production software, such as DaVinci Resolve, AVID, and Adobe Premiere, shown as Figure 1, working as a file classification reference. The Slate board could be previewed in the spotlight of this take. The script of the continuity report could be noted by the Digital Imaging Technician (DIT), who is in charge of the management and verification of raw videos. Otherwise, the editing assistant will import this information into the metadata via the metadata editing workspace in post-production software manually. The works mentioned above are important in the filmmaking process but still need to be optimized due to their inefficiency, time-consuming, non-standardization, and experience-dependent.

Considering manually annotated information is highly semantic. Inspired by computer-vision-based classification and recognition tasks, which have been proved to be practical and effective, we would like to embed semantic understanding methods with automatic metadata generation in order to improve the efficiency of the film raw video retrieval. Semantic label extraction and video caption generation could be used to construct the highly-integrated metadata, which could meet the requirements of post-production artists and users. Taking the use of intelligent and automatic methods, the sematic label and value could be extracted and stored into the metadata structure, and re-construct the comprehensive metadata file for video search and retrieval inside the post-production software. This approach can be based embedded into the existing film production process.

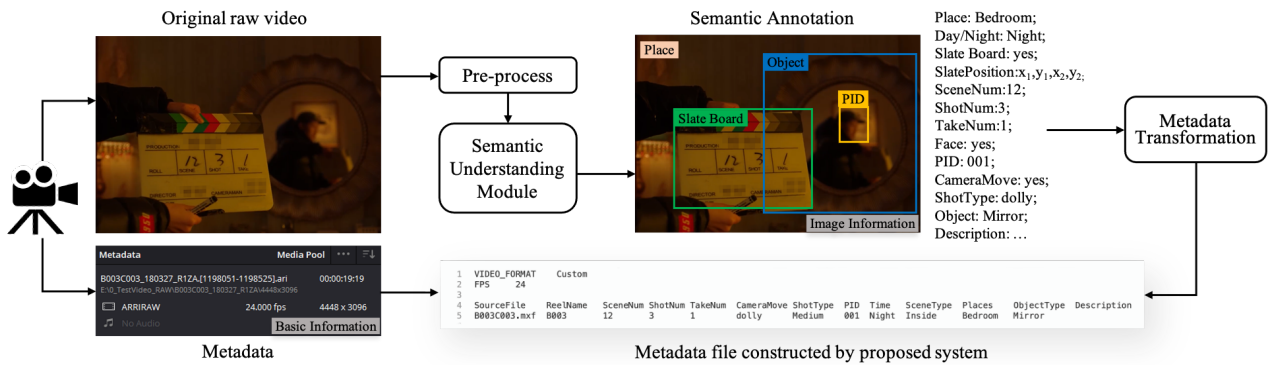

Figure 2: Concept Figure. Our system integrates basic information, recorded by a camera, and image information, extracted as semantic annotations, into metadata file for raw video search and retrieval in the film post-production process.

In this work, we propose a metadata extraction method with semantic understanding, shown as Figure 2, including semantic label annotation and user-demand-oriented information fusion, for video retrieval in film production. We also collected a film raw video retrieval benchmark dataset to analyse the efficiency of the proposed system. The system could extract, transform, and import different metadata structures, adapting to the film-production industrial software, so that post-production artists could choose the format they use. The proposed method improves the efficiency of retrieval, which could be essential and convenient for collaborative work.

## 2 RELATED WORK

Video retrieval is widely used in film recommendation and search for stream media platforms. As usual, when we search for a film, we type in keywords which is related to the film information, such as names of film, directors, and actors. This kind of information, which we named instance information, comes from the crew, script, content, and abstract of the film, given by film producers and distributors. However, the films raw video search and retrieval task in film production is different from the normal video retrieval to some extent. In our work, we focus on the video retrieval in the filmmaking process, which pays more attention to image information, including the shot, visual, and content information, such as scene number, shot number, take number, camera movement, shot scale, actor, action, places, etc. These image information could be extracted by semantic understanding methods to generate comprehensive metadata for search and retrieval. Inspired by deep visual-semantic embedding and image captioning, we would like to employ methods in semantic understanding, video retrieval, and film asset management for reference to address this practical problem.

**Semantic Understanding** for content-based visual information extraction has become an important research topic to discover high-level semantic descriptions within an image. Based on semantic understanding, relevant keywords of a video clip could be extracted through text obtained from video metadata, speech-to-text, and clip summarizies [10]. To align and leverage visual data, a deep visual-semantic learning model could generate representations of image-region-level descriptions [8]. Deep-learning-based methods, such as semantic understanding of visual scenes [20], face localisation [1], textual description [16], and segmentation of images [4], also play a significant role in semantic extraction of visual content and show an exploring strategy for aggregating the context of image information.

**Video Retrieval** is a critical challenging issue to search for a designative video with natural language queries, condensing multi-modal and high-dimension information into a single and compact representation [9]. In the literature, two strategies could be employed for this task, annotation-based and visual-feature-based approaches [14]. The former indexing system takes use of textual descriptions, attribute information, and keywords to represent the video content [17], whereas the latter is focusing on aggregating the frame visual features into a global descriptor, such as deep-features [7] and sentence-level-features [11], which could generate rich dynamic content of video for searching, retrieving, or browsing [18]. Video retrieval methods mentioned above mainly distinguish videos with regional contextual relationships between entities, which is usually used for video recommendation and ranking [6]. However, in view of the fact that raw videos may be shot in the same scene from different angles of the camera, with a strong intra-instance difference and inter-instance similarity, which makes the application of video retrieval methods more difficult to meet the requirement in film production.

**Asset Management** works as a discipline that provides methods and tools for higher efficiency of production and operation. For traditional film asset management in practical, Digital Imaging Technicians lead the effort to build collaborating file management pipeline, including standardized rules of file naming, authorization, and replica, with the help of daily shooting reports, Filemaker software, and a storage server. Nowadays, most state-of-research asset management tools focus on tracking, exploring, and retrieving data to conduct structural knowledge and organized information, offering collaboration and workflow-execution-related operations [5], which provides a new idea for film raw video management. With the development of advanced digital film technologies, such as virtual production, digital double actors, and VFX composition, the film assets grow up in diversity and amount. Human-in-the-loop systems could accelerate repeated workflows by intelligently tracking changes and intermediate results over time [15], which offers the opportunity to realize the user-demand-oriented information fusion. Moreover, industrialization gives more challenges to address the data verification and handover process in an efficient way and support various film industrial post-production software.

## 3 SYSTEM ARCHITECTURE

In this work, we propose a metadata extraction system, which is embedded with semantic understanding and video content analysis, aiming for user-demand-oriented search and raw video retrieve. This system could format the metadata according to the user demand and specialized post-production software, and attach it to the basic information recorded by the camera. In this section, the system architecture and the main methodology modules are presented with the user interface and function introduction as well.

### 3.1 Pre-process

The original raw videos are in RAW formats, such as ARRIRAW, which is developed by ARRI digital film camera manufacturer. RAW format video is recorded and created by a digital camera, without any processing of light information transmitted to the sensor to preserve the highest quality with the minimum compression. These original videos are usually managed, transferred, stored, and copied by DIT from the camera to the workstation directly in the film production process. In order to use the original video in RAW, named raw video, as our input, there are two main pre-processing steps in our system, de-Bayer and image transformation.

**De-Bayer**. Raw video can be converted into RGB video only after de-Bayer processing, which is the process of transforming the original single-channel image sequence, which represents the original Bayer array recorded by the camera sensor, into the RGB three-channel image sequence. Meanwhile, the metadata is generated together in the RAW format.

**Image Transformation**. When shooting a film, raw videos are usually created in LOG, which is a video recording mode that applies the logarithmic function to the exposure curve, resulting in preserving the details of highlights and shadows. Different camera manufacturers developed different LOG modes, such as C-log, N-log, and S-log, from Canon, Nikon, and Sony respectively. Raw videos in LOG visually look grayer than normal color. As we all know, the computer-vision-based semantic understanding models have a certain color sensitivity, so color gamut conversion is needed before the semantic extraction. By applying look-up-table (LUT) to the raw video, the color of the image sequence could be converted into RGB colorspace, which is called primary color matching in film post-production. Because raw videos are in extremely high resolution(4K or 8K) and frame rate (60fps, or even 120fps), it is also necessary to reduce the resolution, down-sample frames, encode, and compress in pre-processing.

### 3.2 Semantic Annotation

After interviewing post-production directors, DITs, script supervisors, and editing assistants, we summarize the limitations of traditional video management methods in film production. Considering a combination of basic, instance, and image information, we select several semantic information for experimental implementation. It should be mentioned that the labels introduced in this system could be extended and amended in terms of user demands. To validate the availability of this system, we build the semantic understanding module with 4 main submodules, named Slate Detection, CameraMove Recognition, Actor Detection, and Scene and Object Recognition, to extract 10 classes of annotations, shown in Table 1.

Table 1: The metadata labels and modules in our system.

| Metadata Class | | Values | Module |
|---|---|---|---|
| Slate | SceneNum | Int | Slate Detection |
| | ShotNum | Int | |
| | TakeNum | Int | |
| Camera | CameraMove | pedestal, dolly, truck, tilt, pan, zoom, etc. | CameraMove Recognition |
| Shot | ShotType | full, medium-full, medium, close, close-up, etc. | Actor Detection |
| Actor | Actor PID | Actor Name | |
| Scene | Time | Day, Night | Scene and Object Recognition |
| | SceneType | Inside, Outside | |
| | Places | Category Name | |
| Object | ObjectType | Category Name | |

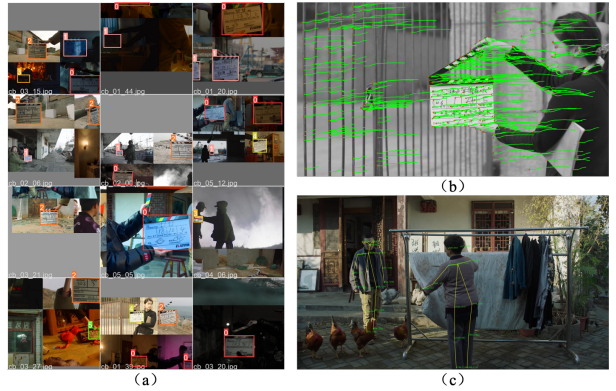

Figure 3: Diagram of slate detection, camera movement recognition, and shot scale recognition in our framework.

#### 3.2.1 Slate Detection

When we shoot a film, each take of the raw video contains a slating action in front of the video, which could offer us helpful information, such as scene number, shot number, take number, director name, cameraman namer, etc. In our work, the slate board detection module, including slate board localization, region segmentation, and OCR, is shown in Figure 3(a). In this system, we retrained the YOLOv5 model [12], extracting feature points with the help of ORB method [13] to align the template image, following with OCR to recognize handwritten digits and words. This module is used to extract the recorded contents and transform them into the data of the corresponding metadata label.

#### 3.2.2 CameraMove Recognition

The main idea of camera movement recognition is to generate the sparse optical flow trajectory of the frame sequence and compare it with patterns of typical camera movement. We extract features in each frame and predict the upcoming points in the next frame to obtain the feature point trajectory. the shown as Figure 3(b). To explain it more clearly, we choose several representative camera movements as references. When the camera pans or tilts, the image transforms in Euclidean transformation. We could reduce noise corner points with RANSAC [3] and add up the trajectory to determine whether its pattern matches the Euclidean transformation matrix. When the camera dollys, the corner points conform perspectively. So that calculating angles between trajectories could be utilized to represent its trajectory pattern. When the cameraman shot handheld, the trajectory shows high-frequency irregular jitter, and the direction of the motion vector changes rapidly. The patterns of trajectory will be used to classify the cameramove types.

#### 3.2.3 Actor Detection

This system module includes actor recognition and shot scale recognition. The detection and recognition of actors can use the widely used location and recognition method to obtain character information. In this system, we employ the pre-trained RetinaFace [1] and ArcFace [2] models for face detection and feature extraction. According to the principles in filmmaking visual language, shot scale is related to the proportion of the subject to the frame, for example, a medium shot usually shows half of the person from head to waist. So that we could calculate the proportion of the height of face to the height of frame to estimate the shot scale. When the person is facing away from the camera, we utilize the 3d pose estimation method to predict the height of the actor, which could be used to estimate the shot scale by calculating its proportion to the height of frame, shown as Figure 3(c)

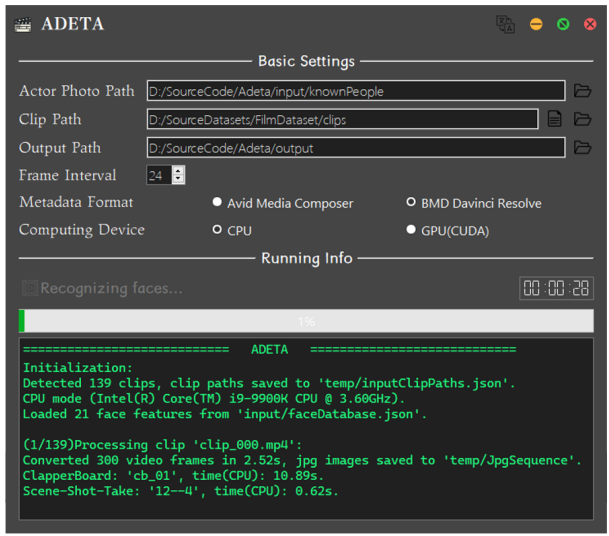

Figure 4: Diagram of our system user interface.

Table 2: Evaluation results of semantic annotation.

|  | SceneNum | ShotNum | TakeNum | Time |
|---|---|---|---|---|
| Accuracy | 0.531 | 0.695 | 0.711 | 0.863 |
|  | PID | ShotScale | CameraMove | SceneType |
| Accuracy | 0.819 | 0.752 | 0.836 | 0.906 |

### 3.2.4 Scene and Object Recognition

For the object and scene recognition module, we need to extract four kinds of labels, including time, scene type, place name, and object category. The normal recognition model can be used for object category recognition, and the scene type and place category can be predicted by the pre-trained places365 scene recognition model [19]. To classify the time, we select day and night binary classification, which is a piece of important information marked for each scene in film scripts. From the perspective of visual color, the night scene seems dark and blue, so that the value of HSV and RGB can be used to classify day and night. Meanwhile, the scene type, inside or outside, should be considered as prior knowledge.

### 3.3 Metadata Transformation

Metadata file format transformation and conversion are integrated into this module. The camera manufacturers have independently-developed metadata file formats as varied as their brand. Meanwhile, post-production softwares also present metadata information in different way. For instance, the metadata file formats of two mainly used post-production software, Avid Media Composer and Davinci Resolve, are in ale and CSV respectively. According to the camera manufacturer and software offered metadata deconstruction APIs, our system could read and extract metadata information, after importing the raw videos into the software media workspace.

Our system fuses user-demand-oriented information into the final metadata information, where allows users to choose the labels they need and the file format that their post-production software could read. The user interface is shown as Figure 4. After parsing the basic information from the original metadata, we will add up the extracted metadata labels or re-construct a new metadata file to extend the diversity of semantic descriptions. All the metadata information can be automatically loaded for classify and index corresponding videos in video management and retrieval.

## 4 IMPLEMENTATION

For validating the proposed strategy, we conducted experiments to evaluate the efficiency of our system and the performance of generated metadata. The experimental configuration is an integration of Slate Detection, CameraMove Recognition, Actor Detection, and Scene and Object Recognition modules, with a graphical user interface. To quantitatively evaluate our system, we built two benchmark datasets, named Film Raw Video Semantic Annotation Dataset (Film-RVSAD) and Slate Board Template Dataset (SBTD) for cinematography-related information extraction with metadata semantic labels.

### 4.1 Evaluation

Film-RVSAD is used to validate the effectiveness of our system. We collected the real film original raw videos and annotated them with semantic information, shown as the upper line in Figure 5. Considering the copyright of raw videos, we were not able to collect a large-scale dataset to retrain models for semantic understanding. In this paper, we build a specific dataset to showcase the effectiveness of the proposed architecture. The performance of our system is shown in Table 2, which is the experimental results by employing pre-trained models and tools introduced above.

In SBTD, we collected six kinds of slate board images that appeared in Film-RVSAD. As shown in Figure 6 (a), all the slate board images were clear and the semantic regions in each image are annotated with the position coordinates of their segments. When training the slate board detection model, we applied image enhancement as well. The detection accuracy are shown in Figure 6 (b).

### 4.2 Results

We present the ability of our proposed system to generate the specific semantic annotations in metadata, shown in Figure 5. Compared with the ground truth, given in the upper line of Figure 5, the proposed system could efficiently extract the required semantic information. As the output of our system, the metadata file can be loaded into post-production software and edited in the metadata workspace as well.

The work we presented in this paper is still in the early stage. The system we designed offers an intelligent data management method for film production, which can effectively improve the efficiency of large-scale raw video search and retrieval in the filmmaking process. However, it is not comprehensive enough. It is necessary to extend and improve the diversity of semantic information and system modules. The accuracy in quantitative evaluation depends on the performance of the selected pre-trained model. However, the benchmark dataset we collected is not rich enough to retrain all the models to improve the accuracy.

With the development of virtual production and advanced VFX technology, multi-stand shooting, multi-scene transition, and collaborative filmmaking workflow present more challenges. Metadata needs to be as detailed and accurate as possible to improve the efficiency of storage, retrieval, browsing, and delivery. In the future, we will continue to conquer and find out solutions for more complex and diverse data retrieval. This paper aims to provide a new application scenario, using the applicable semantic analysis model to effectively browse raw videos, so that we could help post-production artists and producers improve their efficiency.

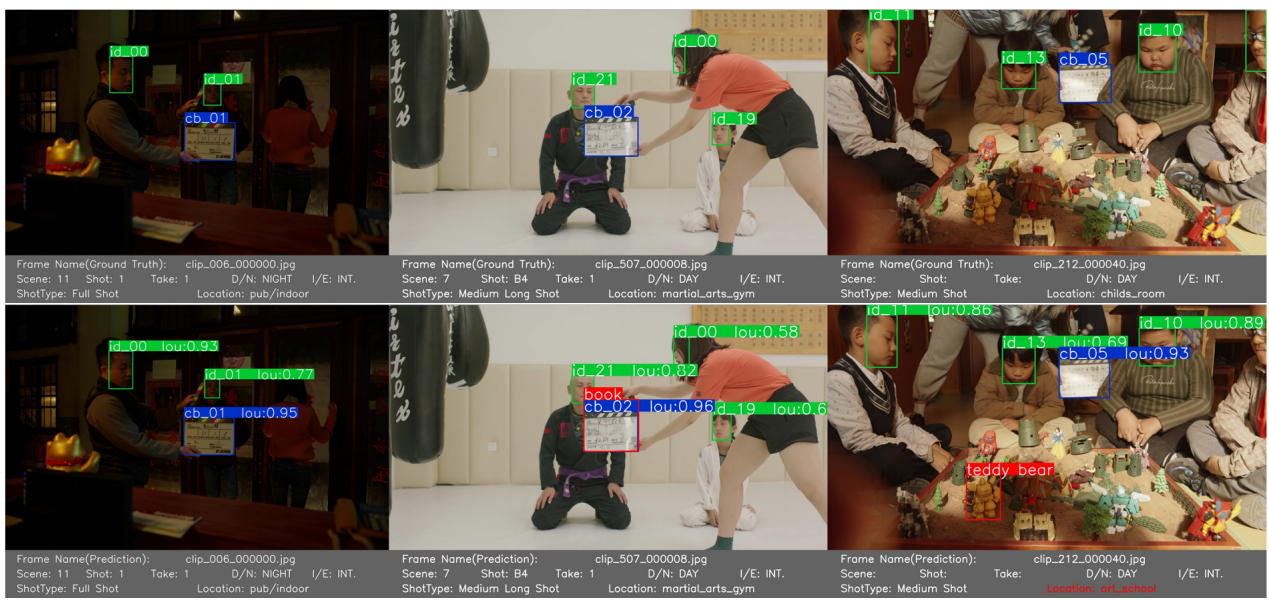

Figure 5: Region predictions and semantic annotations on Film-RVSAD. We use our method to generate metadata for video search and retrieval. Examples in the upper line are ground-truth in Film-RVSAD, and examples in the bottom line are results from our proposed system.

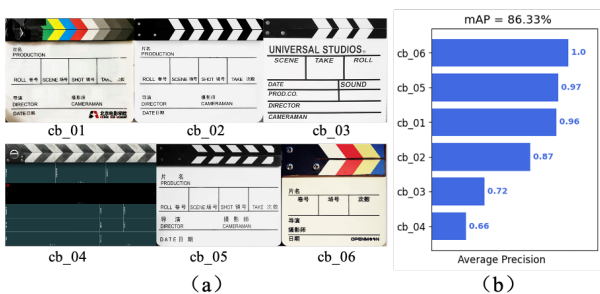

Figure 6: The Slate Board Template Dataset samples and the performance of slate board detection module.

## 5 CONCLUSION

We introduced an application scenario of browsing the massive large-scale raw videos with the help of metadata. In this paper, we propose an automated method with semantic understanding for raw video search and retrieval. Our system utilizes deep visual-semantic models to extract cinematography-related image information, which can be integrated into metadata labels. We conducted experiments to validate the effectiveness of our system. We showed that the proposed system provides an applicable metadata extraction strategy, generating semantic annotations in terms of user demand, to improve the efficiency of video retrieval in film production.

## 6 ACKNOWLEDGEMENT

This work was supported by Beijing Natural Science Foundation (No. 4214073) and the National Social Science Fundation Art Project (No. 20BC040). Special thanks to all the cast and crew from *Nian Nian*, *The Deep Blue*, *Spring in the Hood*, *MiLaZhiGe*, *YueShi*, *TianMingYiFuChouJi* for generously offering original raw videos to support our research.